# Nanoscale visualization and spectral fingerprints of the charge order in ScV$_6$Sn$_6$ distinct from other kagome metals

Siyu Cheng[1,*], Zheng Ren[2,*], Hong Li[1], Ji Seop Oh[3,2], Hengxin Tan[4], Ganesh Pokharel[5], Jonathan M. DeStefano[6], Elliott Rosenberg[6], Yucheng Guo[2], Yichen Zhang[2], Ziqin Yue[2,13], Yongbin Lee[7], Sergey Gorovikov[8], Marta Zonno[8], Makoto Hashimoto[9], Donghui Lu[9], Liqin Ke[7], Federico Mazzola[10,11], Junichiro Kono[2], R. J. Birgeneau[3,12], Jiun-Haw Chu[6], Stephen D. Wilson[5], Ziqiang Wang[1], Binghai Yan[4], Ming Yi[2,&] and Ilija Zeljkovic[1,^]

[1]Department of Physics, Boston College, Chestnut Hill, MA 02467, USA

[2]Department of Physics and Astronomy, Rice University, Houston, TX 77005, USA

[3]Department of Physics, University of California, Berkeley, CA 94720, USA

[4]Department of Condensed Matter Physics, Weizmann Institute of Science, Rehovot, Israel

[5]Materials Department, University of California Santa Barbara, Santa Barbara, California 93106, USA

[6]Department of Physics, University of Washington, Seattle, Washington 98195, USA

[7]Ames Laboratory, Ames, IA, 50011, USA

[8]Canadian Light Source, Inc., 44 Innovation Boulevard, Saskatoon SK S7N 2V3, Canada

[9]Stanford Synchrotron Radiation Lightsource, SLAC National Accelerator Laboratory, Menlo Park, California 94025, USA

[10]Istituto Officina dei Materiali (IOM)–CNR, Laboratorio TASC, Area Science Park, S.S.14, km 163.5, I-34149 Trieste, Italy

[11] Department of Molecular Sciences and Nanosystems, Ca' Foscari University of Venice, 30172 Venice, Italy

[12]Materials Science Division, Lawrence Berkeley National Laboratory, Berkeley, CA 94720, USA

[13]Applied Physics Graduate Program, Smalley-Curl Institute, Rice University, Houston, TX 77005, USA

*contributed equally

Corresponding authors: my32@rice.edu[&] and ilija.zeljkovic@bc.edu[^]

## Abstract

**Charge density waves (CDWs) have been tied to a number of unusual phenomena in kagome metals, including rotation symmetry breaking, time-reversal symmetry breaking and superconductivity. The majority of the experiments in kagome metals thus far have focused on the CDW states in $AV_3Sb_5$ ($A$=K, Cs, Rb) and FeGe, characterized by the 2$a_0$ by 2$a_0$ CDW period in the kagome plane. Recently, a bulk CDW phase ($T_{CDW}$ ≈ 92 K) with an entirely different wave length and orientation has been reported in ScV$_6$Sn$_6$, as the first realization of a CDW state in the broad $RM_6X_6$ crystal structure. Here, using a combination of scanning tunneling microscopy/spectroscopy and angle-resolved photoemission spectroscopy, we reveal the microscopic structure and the spectroscopic signatures of this charge ordering phase in ScV$_6$Sn$_6$. Differential conductance d$I$/d$V$ spectra show a partial gap opening in the density-of-states of about 20 meV at the Fermi level. This is much smaller than the spectral gaps observed in $AV_3Sb_5$ and FeGe despite the comparable $T_{CDW}$ temperatures in these systems, suggesting substantially weaker coupling strength in ScV$_6$Sn$_6$. Surprisingly, despite the three-dimensional bulk**

nature of the charge order, we find that the charge modulation is only observed on the kagome termination. Temperature-dependent band structure evolution suggests a modulation of the surface states as a consequence of the emergent charge order, with an abrupt spectral weight shift below $T_{CDW}$ consistent with the first-order phase transition. The similarity of the electronic band structures of ScV$_6$Sn$_6$ and TbV$_6$Sn$_6$ (where the charge ordering is absent), together with the first-principle calculations, suggests that charge ordering in ScV$_6$Sn$_6$ may not be primarily electronically driven. Interestingly, in sharp contrast to the CDW state of cousin $A$V$_3$Sb$_5$, we find no evidence supporting rotation symmetry breaking. Our results reveal a distinctive nature of the charge ordering phase in ScV$_6$Sn$_6$ in comparison to other kagome metals.

**Introduction**

The kagome lattice, a pattern of tessellated hexagons connected by small corner-sharing triangles, emerged as a versatile platform for exploring a variety of novel quantum phases of matter. Due to the geometric frustration intrinsic to the kagome lattice, layered kagome materials are characterized by a characteristic electronic band structure consisting of Dirac fermions, flat bands and Van Hove singularities (VHSs) [1–5]. This prototypical band structure can be intertwined with a rich array of exotic electronic instabilities, which have been theoretically explored and experimentally realized in several families of kagome metals thus far. For example, Fe- [6–15], Mn- [16–21] and Co- [22–25] based kagome magnets yielded the realization of topological flat bands [11,23,26], Dirac and Weyl fermions [6,9,12,22,24,27] and Fermi arcs [22,24]; a non-magnetic V-based $A$V$_3$Sb$_5$ (A=Cs, K, Rb) kagome metal family [28–30] on the other hand attracted a large interest [31–41] due to the emergence of superconductivity and various symmetry-breaking states potentially in connection to loop current orders [42].

The recently discovered bilayer kagome metals in the $R$V$_6$Sn$_6$ structure ("166" family), where $R$ stands for a rare earth ion, offer a new tunable platform to investigate Fermi surface instabilities of the kagome lattice [43–48]. Similarly to that in $A$V$_3$Sb$_5$, vanadium atoms that comprise the kagome layers in $R$V$_6$Sn$_6$ remain non-magnetic [44–46,49], but magnetism can still be selectively tuned by the choice of the rare earth element $R$ [49–51]. Out of the wide array of kagome metals in the $R$V$_6$Sn$_6$ structure, ScV$_6$Sn$_6$ presents a unique platform where a charge density wave (CDW) state was reported below $T_{CDW} \approx 92$ K [46]. Similar to the bulk CDW phase in $A$V$_3$Sb$_5$ that is three-dimensional, this charge ordering (CO) phase in ScV$_6$Sn$_6$ is also three-dimensional but with a different wave vector $\mathbf{Q}^*$= (1/3, 1/3, 1/3). Understanding the formation of the CO and how it compares to the more intensely investigated CDW in $A$V$_3$Sb$_5$ has been of high interest, but very little is known about the origin and the spectroscopic fingerprint of the charge order in ScV$_6$Sn$_6$.

Here we use a combination of scanning tunneling microscopy/spectroscopy (STM/S) and angle-resolved photoemission spectroscopy (ARPES) to investigate the electronic structure and the CO formation in ScV$_6$Sn$_6$. STM topographs reveal a $\sqrt{3} \times \sqrt{3}$ $R30°$ electronic superstructure, which corresponds to the in-plane component of the bulk CO wavevector identified in diffraction measurements. We visualize the three-dimensional nature of the CO by imaging the expected phase offset across different terraces. Interestingly, the microscopic signature of the CO at the surface is termination-dependent – while a small, partial spectral gap at the Fermi level and the CO peaks are observed on the kagome termination, both are notably absent on the Sn termination. In contrast to the C$_2$-symmetric CDW state in $A$V$_3$Sb$_5$, we discover that the rotational symmetry in the CO state of ScV$_6$Sn$_6$ appears to be preserved. Our low-temperatures ARPES measurements reveal that the electronic structures of ScV$_6$Sn$_6$ (with the CO) share much in common with that of TbV$_6$Sn$_6$ (without the CO). Lastly, we uncover the temperature evolution of the electronic bands in the vicinity of the K-point, which is consistent with the modulation of the surface bands in first-principle calculations as a consequence of the CO.

**Results**

*Microscopic visualization of the CO structure*

Crystal structure of ScV$_6$Sn$_6$ is composed of layers of V atoms arranged on a kagome network, each stacked between a ScSn$^3$ layer and a Sn$^1$-Sn$^2$-Sn$^1$ trilayer (Fig. 1a,b). We cleave bulk single crystals of ScV$_6$Sn$_6$ in ultra-high vacuum to expose a pristine surface before measurements (Methods). Related materials in the same crystal structure tend to cleave along the *c*-axis to reveal *ab*-plane surface terminations [19,20,27,43,45,47]. This is consistent with the observed surface structure of UHV-cleaved ScV$_6$Sn$_6$ – STM topographs exhibit a hexagonal lattice (Fig. 1e-h) and the occasionally observed steps are an integer number of *c*-axis unit cell heights (Fig. 1c, Fig. 3). STM topographs show two types of surface morphologies, both with a hexagonal lattice (Fig. 1e,g, Supplementary Figure 1, Supplementary Figure 2). We identify the termination in Fig. 1e as the Sn$^2$ termination due to the clearly resolved individual atoms forming a hexagonal lattice (top inset in Fig. 1e) similar to the Sn$^2$ surface imaged in YMn$_6$Sn$_6$ [20]. The surface termination in Fig. 1g is likely the kagome layer, as we deem the other two possibilities, Sn$^1$ and ScSn$^3$, to be unlikely: the Sn$^1$ termination in YMn$_6$Sn$_6$ showed a triangular atomic structure [20] and the *Re*Sn$^3$ surface appears to have a tendency to reconstruct into stripe-like features as reported in TbMn$_6$Sn$_6$ [27].

In addition to the atomic Bragg peaks $\mathbf{Q}^i_{Bragg}$ (*i*=*a*, *b* or *c*), Fourier transforms (FTs) of STM topographs of the V kagome termination show six additional superlattice peaks (Fig. 1h). These FT peaks are positioned exactly along each Γ-K direction at $\mathbf{Q}^* = (\mathbf{Q}^i_{Bragg}+\mathbf{Q}^j_{Bragg})/3$ (*i, j* =*a*, *b* or *c*) and are symmetric with respect to the center of the FT. They correspond to the staggered intensity modulations seen in real-space in STM topographs (inset in Fig. 1g). To investigate the origin of these modulations, we examine the energy dependence of the associated wave vectors in the FTs of differential conductance d*I*/d*V*(**r**,*V*) maps. The FT peaks are discernable in a wide range of energies *e·V* and the peak positions in momentum-transfer space are independent of the bias *V*. Moreover, the peak position is exactly consistent with the in-plane component of the newly discovered CO phase detected in diffraction experiments [46]. Therefore, the charge modulations observed in the STM topograph in Fig. 1g represent a microscopic visualization of the bulk CO state at the surface of this system. Differential conductance d*I*/d*V* spectra on the kagome surface termination show a gap-like feature, with a partial suppression in the density-of-states of about 20 meV, approximately symmetric with respect to the Fermi level (Fig. 2e). As the CO is the only known ordered state in this system, we deem that the gap is likely related to the emergent CO.

To gain insight into the three-dimensional nature of the order revealed to have a (1/3, 1/3, 1/3) wave vector from diffraction measurements [46], we study the charge modulations across a step edge (Fig. 3). Our theoretical calculations suggest that structural distortions from the CO create an effective (2π/3, 2π/3) phase shift between the adjacent kagome bilayers (Fig. 3b). As the structural distortion is in principle intimately tied to the charge modulation pattern observed in STM measurements, one would also expect to observe an apparent charge modulation phase shift between adjacent atomic terraces. To explore this, we study in detail an STM topograph between two different terraces (Fig. 3c). Both terraces are terminated by the vanadium layer and show the same CDW pattern that can be clearly resolved in real space (Fig. 3c). By tracking the positions of the charge modulation peaks on the upper terraces, we observe a phase shift of the equivalent modulations peaks on the lower terrace (Fig. 3c). This is consistent with the expected bulk CO structure (Fig. 3b), thus providing the first microscopic insight into the three-dimensional nature of the CO that persists at the surface of the material.

*Absence of rotational symmetry breaking in the CO state of ScV$_6$Sn$_6$*

In the CDW phase of $A$V$_3$Sb$_5$, an array of experimental techniques observed signatures of in-plane rotational symmetry breaking, which first onsets at the $T_{CDW}$ temperature [38,52,53] and reduces the in-plane rotational symmetry from six-fold to two-fold [31,32,36,38,52–54]. We proceed to explore whether the rotational symmetry of the kagome bilayer in ScV$_6$Sn$_6$ is also broken. For pedagogical purposes, we first discuss the experiments on KV$_3$Sb$_5$. Similarly to the analysis done in Refs. [36,52,54], we compare the CDW amplitudes along the three inequivalent directions as a function of energy (Fig. 4c). A noticeable difference in the CDW amplitude dispersions is clearly observed between different peaks, with one direction being markedly different from the other two that are nearly the same. This FT peak anisotropy is also reflected in the unidirectionality of the pattern in d$I$/d$V$(**r**,$V$) maps (Fig. 4d). In ScV$_6$Sn$_6$ however, we find that the shape of the CO amplitude dispersions versus energy along the three directions is nearly identical (Fig. 4a). We note that tiny differences between the three curves reflect the measurement and analysis uncertainty, as well as the inevitable small STM tip anisotropy. The approximately rotationally symmetric CO signature can also be seen in representative d$I$/d$V$(**r**,$V$) maps (Fig. 4b). These measurements suggest the absence of rotation symmetry breaking in the CO state of ScV$_6$Sn$_6$.

*Spectroscopic investigation of the nature of the CO*

Having identified the real-space signatures of the CO in ScV$_6$Sn$_6$, we next turn to the momentum-space evidence. ARPES has been an extremely useful tool in determining the signature kagome bands [14,41,55–58], the CDW gap modulation and the band reconstruction associated with CDW order [57–59] in $A$V$_3$Sb$_5$ and FeGe where CDWs have been reported thus far. We first show the measured Fermi surface (FS) of ScV$_6$Sn$_6$ obtained on the kagome termination (see Supplementary Note 1, Supplementary Figure 3). The FS consists of segments connecting the M points of the Brillouin zone (BZ) (Fig. 5a), similar to that of other kagome metals such as $A$V$_3$Sb$_5$ [29,41,57,59] and FeGe [14,58]. From the dispersions measured along the high symmetry direction Γ-K-M, we observe a quadratic band centered at Γ, with the band bottom located at about 1 eV below the Fermi level, $E_F$. This band extends to the M point in the form of a VHS near $E_F$ (Fig. 5c). This is consistent with the typical band structure associated with the kagome lattice, and also matches well with the band structure calculated by the density-functional theory (DFT) (Fig. 5d), suggesting weak electron-electron correlation effects. Since both ScV$_6$Sn$_6$ and TbV$_6$Sn$_6$ share the same V kagome layer, we proceed to compare the band structures of the two systems. Both the FS and band dispersions show a striking similarity. In particular, the locations of the VHS at M are very close to $E_F$ in both systems, with that in ScV$_6$Sn$_6$ slightly below $E_F$ and that in TbV$_6$Sn$_6$ slightly above $E_F$ (Fig. 5c,g). The small difference in VHS positions is also consistent with the DFT calculations (Fig. 5d,h).

To investigate the effect of the CO on the electronic structure of ScV$_6$Sn$_6$, we perform temperature-dependent measurements of dispersions along the Γ-K-M direction across the transition T* ≈ 92 K [46]. To reveal the band reconstruction, we present a series of normalized intensity plots, showing the subtracted difference between the cuts taken at selected temperatures and that taken at 120 K > T* (Fig. 6b) (see Supplementary Figure 4). The most substantial difference appears in the vicinity of the K points near $E_F$, seen as patches of blue on these color maps, and disappears above about 100 K (Fig. 6b). This difference can also be clearly seen by comparing the energy distribution curves (EDCs) taken at 20 K and 120 K in the region across the K point, where we discover a prominent spectral weight shift towards $E_F$ on the side of the K point closer to the M point (upper half of Fig. 6c). The temperature evolution of the spectral weight change can also be seen in the collapsed stack of integrated EDCs in this momentum-space region (Fig. 6d). To quantitatively determine the temperature dependence of the spectral weight shift, we plot the peak area of this EDC integrated between -0.2 V and $E_F$ (Fig. 6d,e). The peak area decreases as the temperature increases, with an abrupt drop across T* (Fig. 6e), consistent with the transition observed. The abrupt drop of the peak area is compatible with the first-order nature of the phase transition [46].

To understand the origin of the band reconstruction near the K points, we perform DFT calculations for the CDW state. We first present the DFT calculations for the bulk bands along Γ-K-M (Fig. 6f). Note that the CDW band structure has been unfolded back into the pristine BZ for ease of comparison. Comparing the pristine and CDW calculations, we observe little difference near the K point. This can be understood as the three-dimensional CDW wavevector (1/3, 1/3, 1/3) corresponds to a reduction of the pristine BZ to a smaller one, where the pristine K and K'$_{1/3}$ (the point at 1/3 along K'-H') points are now equivalent in the reconstructed BZs (Fig. 6h). However, DFT calculations for bulk states predict no states in the vicinity of H point near $E_F$ in the pristine phase (Supplementary Figure 5, Methods), which would lead to negligible folding between K and K'$_{1/3}$ points. Hence, the band reconstruction we observe near the K point is unlikely to be directly due to bulk band folding. However, when the surface states are subject to the CDW folding potential, they would fulfill the in-plane folding condition $\mathbf{Q}^*_{in-plane} = (\mathbf{Q}^a_{Bragg}+\mathbf{Q}^b_{Bragg})/3$, folding states between the K' and K points. We carried out DFT calculations for the surface states on the kagome termination under the bulk CDW wavevector (1/3, 1/3, 1/3). Indeed, folded bands and gap openings appear near the K point (Fig. 6g). Specifically, immediately below $E_F$ at K, a band crossing opens a gap and the upper branch is pushed towards $E_F$ (blue arrow in Fig. 6g). This is consistent with the spectral weight shift towards $E_F$ we observe at the K point (Fig. 6i).

**Discussion**

Our experiments provide the first microscopic visualization of the CO phase in ScV$_6$Sn$_6$ and its spectroscopic fingerprints. The spectral gap observed in d$I$/d$V$ spectra is about 20 meV, which is much smaller than the CDW gap observed in d$I$/d$V$ spectra of AV$_3$Sb$_5$ and FeGe of about 40-50 meV [31,35,36,60], despite the comparable T$_{CDW}$ temperatures in the two systems. This indicates a substantially weaker coupling in the CO state of ScV$_6$Sn$_6$ compared to AV$_3$Sb$_5$. While charge modulations are clearly observed on the kagome termination, we note that Sn$^2$ surface termination does not show the CO in either the topograph or d$I$/d$V$(**r**,$V$) maps near $E_F$. Correspondingly, the small spectral gap in d$I$/d$V$ spectra on the kagome termination is notably absent on the Sn termination (Fig. 2e,f). These observations should be highly relevant for other surface sensitive studies, which should take into account the unusual termination dependence of the bulk order in this system. For comparison, the 2 x 2 CDW state in AV$_3$Sb$_5$ can be detected in STM/S measurements of both surface terminations imaged [31,32,35,36].

Our results also suggest that while the CO in ScV$_6$Sn$_6$ breaks the translational symmetry of the lattice, the rotational symmetry of the kagome bilayer is preserved. This is again in contrast to AV$_3$Sb$_5$, where rotation symmetry breaking occurs concomitant with the CDW onset[38,52,53]. It is conceivable that rotation symmetry breaking in AV$_3$Sb$_5$ is related to the unusual orbital picture as recently evidenced in isostructural CsTi$_3$Bi$_5$ [61,62].

Finally, we comment on the role of VHSs in ScV$_6$Sn$_6$ in comparison to other kagome metals. The VHSs have now been observed at the M points of the BZ near E$_F$ in both ScV$_6$Sn$_6$ and TbV$_6$Sn$_6$, as well as AV$_3$Sb$_5$ and FeGe. In AV$_3$Sb$_5$ and FeGe, it is natural to relate the 2 x 2 CDW order to the FS nesting as the wavevector in principle fulfills the nesting condition. However, in the present case of ScV$_6$Sn$_6$, while the vHSs remain close to E$_F$ at the M point, the CO wavevector is no longer compatible with the same FS nesting wave vectors. Hence, we deem that the CO in ScV$_6$Sn$_6$ may not be primarily electronically driven. This is further supported by the similarities in the fermiology of ScV$_6$Sn$_6$ and TbV$_6$Sn$_6$, and the contrasting presence of the CO in the former and the absence in the latter. In the case of ScV$_6$Sn$_6$, DFT calculations of the phonon spectra also indicate the presence of phonon softening at the K point of the BZ [63]. Interestingly, recent Monte Carlo simulations revealed the importance of the proximity of a Dirac point to E$_F$ in stabilizing the $\sqrt{3} \times \sqrt{3}$ CDW order on a kagome lattice [64], a condition that appears to be satisfied in our system (Fig. 5). In the larger context of kagome metals, our observations suggest that the nesting of VHSs may be a contributor but not the sole driver of the CO, and the combination of the favorable Dirac cone placement near E$_F$ and phonon softening at K could play an important role here. Overall, the charge modulation observed in ScV$_6$Sn$_6$ is likely to be more akin to the lattice-driven structural transition often associated with first order phase transitions rather than the conventional electronically-driven CDW order associated with a second order phase transitions.

In conclusion, we provide both spatial- and momentum-resolved evidence of the unusual CO in ScV$_6$Sn$_6$. We observe a $\sqrt{3} \times \sqrt{3}$ $R30°$ superstructure by STM that corresponds to the in-plane component of the CO wavevector Q* = (1/3, 1/3, 1/3), as well as a phase shift across an atomic step edge that is consistent with the reported 3D CO periodicity along the *c*-axis. We find the rotation symmetry to be preserved in the CO state of ScV$_6$Sn$_6$ in sharp contrast to the C$_2$-symmetric CDW in *A*V$_3$Sb$_5$ (A=Cs, K, Rb) family. In momentum space, we observe the presence of the signature VHSs near the M point of the BZ near E$_F$, yet band reconstruction across T* to occur dominantly near the K points of the BZ, compatible with the folding vector of Q* of surface bands. The similarity of the electronic structure of ScV$_6$Sn$_6$ and TbV$_6$Sn$_6$ where the ordering is absent strongly suggests a non-electronic origin of the CO in ScV$_6$Sn$_6$. Taken all evidence together, our results reveal a distinctive nature of the CO in ScV$_6$Sn$_6$ in comparison to other kagome metals.

**Methods**

*Single crystal growth and characterization.* Single crystals of ScV$_6$Sn$_6$ were grown from Sc (pieces, 99.9%), V (pieces, 99.7%), and Sn (shots, 99.99%) via a flux-based growth technique. The flux mixture of Sc, V and Sn was loaded inside an alumina crucible with a molar ratio of 1:6:40 and then heated at 1125°C for 12 h. Then, the mixture was cooled at a rate of 2°C/hr to 780° C and centrifuged to separate the single crystals from the excess Sn-flux.

*STM experiments*. Single crystals of ScV$_6$Sn$_6$ were cleaved in ultrahigh vacuum and inserted into the STM head at 4.5 K. All STM measurements were taken at about 4.5 K, using home-made electrochemically etched tungsten tips annealed in UHV before STM experiments. STM data were acquired using a customized Unisoku USM1300 STM system. Spectroscopic measurements were made using a standard lock-in technique with 910-Hz frequency and bias excitation as noted in the figure captions.

*Angle-resolved photoemission spectroscopy (ARPES) experiments.* ARPES measurements on ScV$_6$Sn$_6$ were performed at the QMSC beamline of the Canadian Light Source, equipped with a R4000 electron analyzer, and TbV$_6$Sn$_6$ at Stanford Synchrotron Radiation Lightsource (SSRL), Beamline 5-2, equipped with a DA30 electron analyzer. The single crystals were cleaved *in-situ* at 20 K and measured in ultra-high vacuum with a base pressure better than 6x10$^{-11}$ Torr. Energy and angular resolution were set to be better than 20 meV and 0.1°, respectively (Supplementary Figure 6).

*Density-functional theory (DFT) calculations.* All calculations are performed within the DFT implemented in the Vienna ab-initio Simulation Package (VASP) [65]. The generalized gradient approximation as parameterized by Perdew, Burke, and Ernzerhof [66] is employed for the exchange-correlation interaction between electrons. The energy cutoff for the plane wave basis set is 300 eV. A force convergence criteria of 1 meV/Å is used in the structural relaxation. The bulk Brillouin zone of the pristine ScV$_6$Sn$_6$ is sampled with a *k*-mesh of 21×21×10. The surface states of both the pristine phase and the CDW phase are simulated with slabs of 8 kagome layers thick, and obtained by projecting the slab band structure onto the surface unit cell. Spin-orbital coupling is considered for the bulk band structure while not considered in slab calculations.

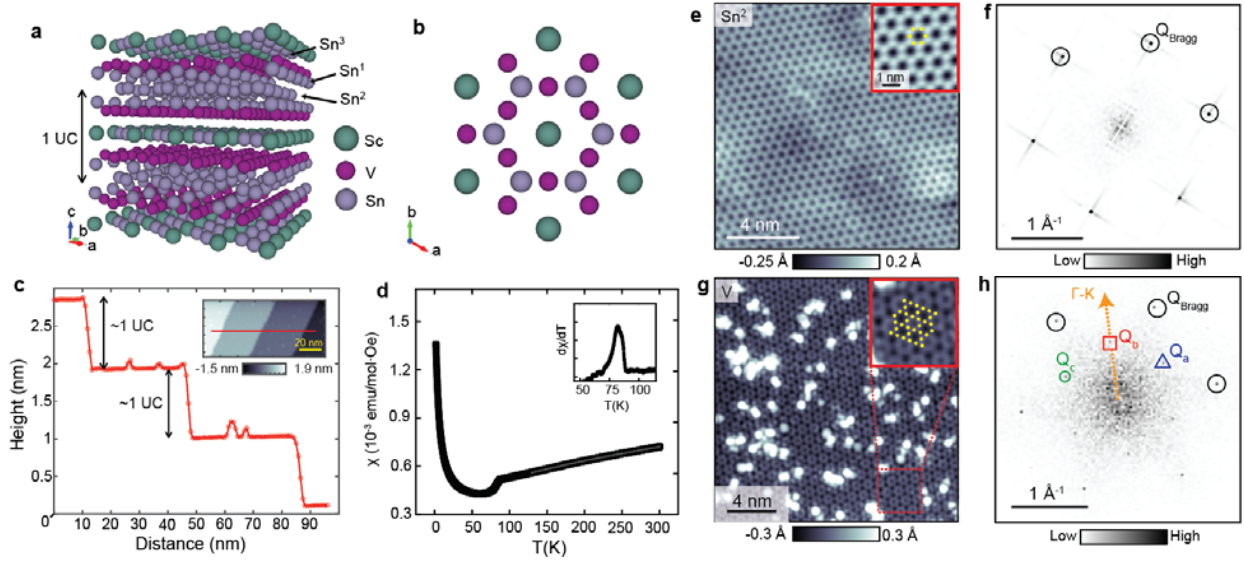

**Figure 1. Crystal structure, magnetization and topographic characterization of ScV$_6$Sn$_6$.** (a) The schematic of the bulk crystal structure of ScV$_6$Sn$_6$, and (b) *ab*-plane surface atomic structure. (c) Topographic linecut showing single steps between consecutive Sn$^2$ terraces taken along the red line in the inset. Inset in (c) shows an STM topograph spanning several consecutive steps. (d) Magnetization measurements (zero-field cooled, then taken at 1 T applied in the *ab*-plane warming up) showing a kink at $T^*$ associated with the bulk transition. STM topographs of (e) Sn$^2$ termination and (g) kagome termination, and (f, h) associated Fourier transforms. Atomic Bragg peaks are circled in green; CO peaks are enclosed in blue circles, red squares and orange triangles. STM setup conditions: (c) $I_{set}$ = 10 pA, $V_{sample}$ = 1V; (e) $I_{set}$ = 400 pA, $V_{sample}$ = 20mV; (g) $I_{set}$ = 200 pA, $V_{sample}$ = 200 mV.

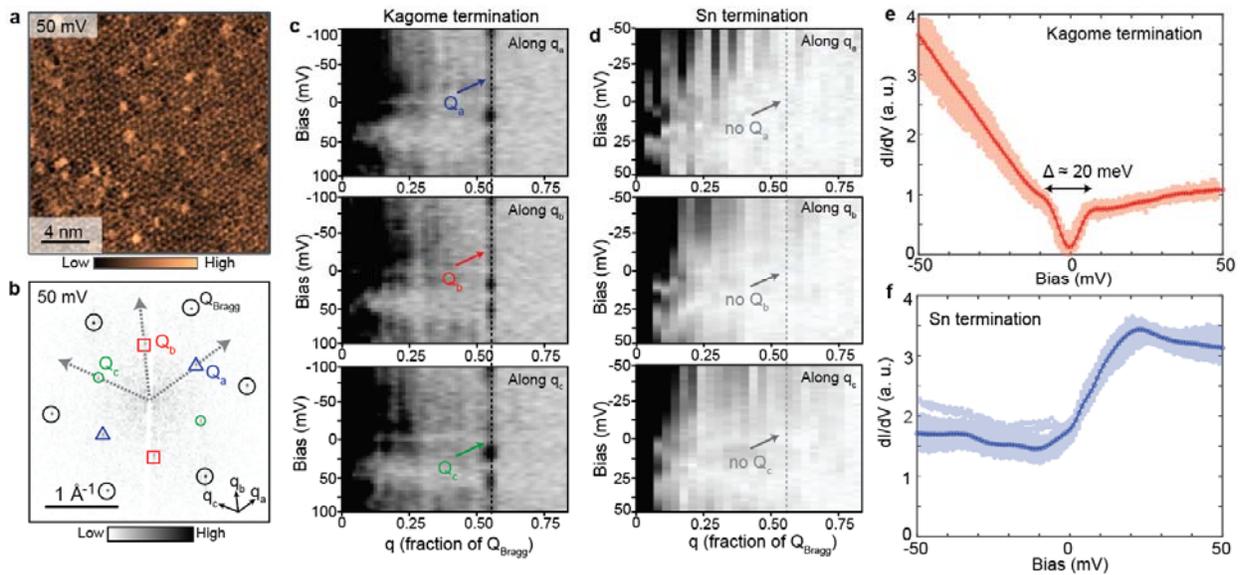

**Figure 2. Termination-dependent scanning tunneling microscopy and spectroscopy of the CO state.** (a) Differential conductance d$I$/d$V$(**r**, $V$=50 mV) map acquired over the kagome termination, and (b) associated Fourier transform (FT). (c) Energy-dependent FT linecuts of d$I$/d$V$(**r**, $V$) maps of the kagome termination, starting at the FT center along the three Γ-K directions (labeled as $q_a$, $q_b$ and $q_c$ in panel (b)). Non-dispersive CO peaks are observed along all three directions. (d) Energy-dependent FT linecuts of d$I$/d$V$(**r**, $V$) maps of the Sn$^2$ termination, starting at the FT center along the three Γ-K directions. No additional peaks are observed at the same momentum-transfer position where the CO peaks are seen on the kagome termination. (e,f) Average d$I$/d$V$ spectra (solid line and circle symbols) over (e) the kagome and (f) Sn$^2$ termination. Diffuse orange (blue) background shows representative d$I$/d$V$ spectra taken over several linecuts in the same field-of-view on kagome (Sn$^2$) surface. STM setup conditions: (a) $I_{set}$ = 50 pA, $V_{sample}$ = 50 mV, $V_{exc}$ = 4 mV; (c) $I_{set}$ = 30 pA, $V_{sample}$ = -100 mV, $V_{exc}$ = 5 mV; (d) $I_{set}$ = 200 pA, $V_{sample}$ = 50 mV, $I_{set}$ = 2 mV; (e) $I_{set}$ = 30 pA, $V_{sample}$ = 50 pA, $V_{exc}$ = 3 mV; (f) $I_{set}$ = 200 pA, $V_{sample}$ = 50 mV, $V_{exc}$ = 1 mV.

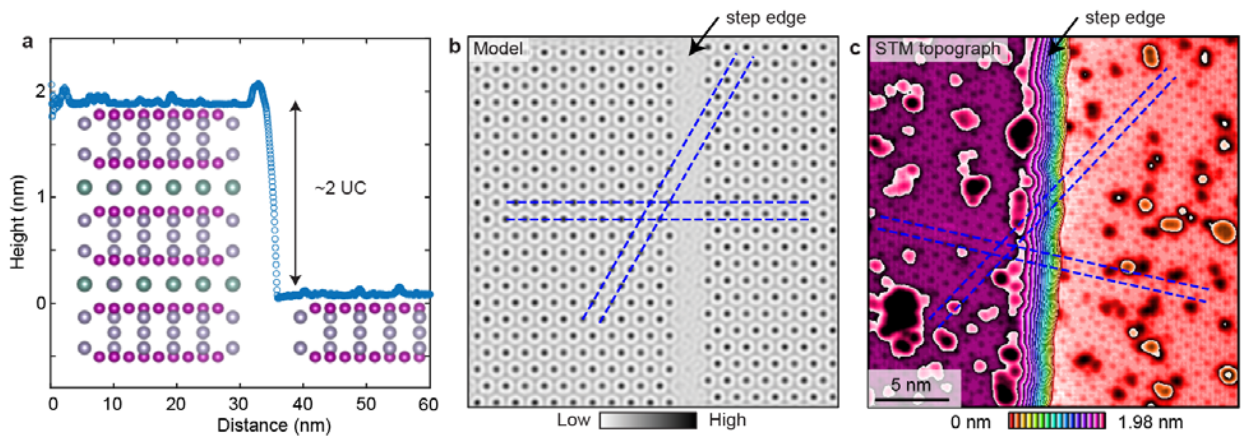

**Figure 3. Visualizing the three-dimensional nature of the CO in ScV$_6$Sn$_6$.** (a) Schematic of the two unit cell step between two kagome-terminated terraces. (b) Simulated CO morphology across the step in (a). Red lines in (b) trace the peaks of the charge modulations on the left half of the image; peaks on the right sides of the image are noticeably offset from the red lines, highlighting the phase shift between the two. (c) STM topograph of the step shown in (a). Similarly to panel (b), yellow lines trace the peaks of the charge modulations on the upper (left) terrace; the charge modulation peaks on the right side (lower terrace) are again visibly away from these lines, demonstrating a phase offset between the two qualitatively consistent with (b). STM setup condition: (a,c) $I_{set}$ = 100 pA, $V_{sample}$ = 1 V.

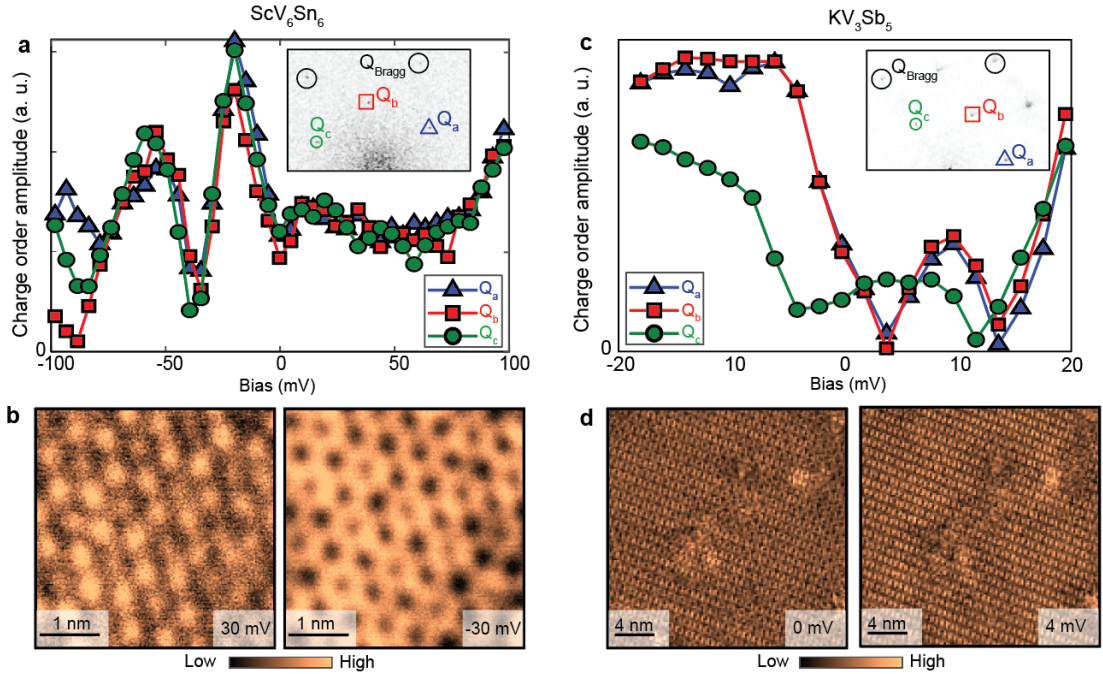

**Figure 4. Scanning tunneling microscopy and spectroscopy investigation of rotation symmetry breaking in ScV$_6$Sn$_6$, and the comparison to KV$_3$Sb$_5$.** (a) Energy-dependent amplitudes of the three inequivalent CDW peaks (marked in the example Fourier transform in the inset) in ScV$_6$Sn$_6$. The overall shape of all three dispersion curves is nearly identical and largely overlaps one another. (b) Zoom-in on representative high-resolution d$I$/d$V$(**r**, $V$) maps showing the approximately rotationally symmetric real-space signature. (c) Energy-dependent amplitudes of the three inequivalent CDW peaks (marked in the example Fourier transform in the inset) in cousin kagome metal KV$_3$Sb$_5$. the dispersion of $Q_c$ amplitude is noticeably different than the other two, $Q_a$ and $Q_b$, that are nearly indistinguishable. (d) Representative d$I$/d$V$(**r**, $V$) maps showing a unidirectional real-space signature that breaks the six-fold symmetry. STM setup conditions: (a) $I_{set}$ = 30 pA, $V_{sample}$ = -100 mV, $V_{exc}$ = 5 mV; (b) $I_{set}$ = 100 pA, $V_{sample}$ = 100 mV, $V_{exc}$ = 2 mV; (c,d) $I_{set}$ = 150 pA, $V_{sample}$ = 10 mV, $V_{exc}$ = 1 mV.

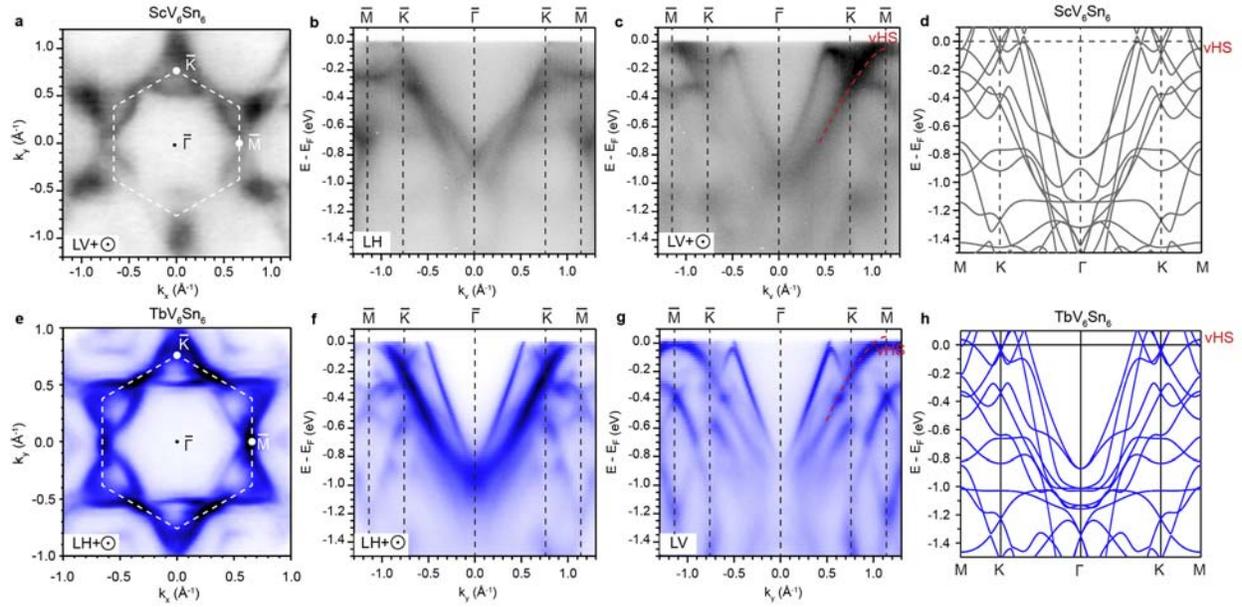

**Figure 5. Comparison between band structure of ScV$_6$Sn$_6$ and TbV$_6$Sn$_6$. a** Fermi surface map of ScV$_6$Sn$_6$ overlaid with the 2D BZ. **b,c** Dispersions acquired along the high-symmetry direction $\bar{\Gamma}$-$\bar{K}$-$\bar{M}$, with the polarization LH and LV+⊙, respectively. Here LH and LV are defined as the in-plane component along the horizontal and vertical directions of the image. When there is an out-of-plane contribution due to experimental setup, we denote it by "+⊙". The red dashed line serves as a guide to the eye for the band forming the VHS near E$_F$. **d** Band structure of ScV$_6$Sn$_6$ in the pristine phase calculated by DFT. **e-h** Data (83 eV photon energy) and DFT calculations of TbV$_6$Sn$_6$ corresponding to each panel of ScV$_6$Sn$_6$ above.

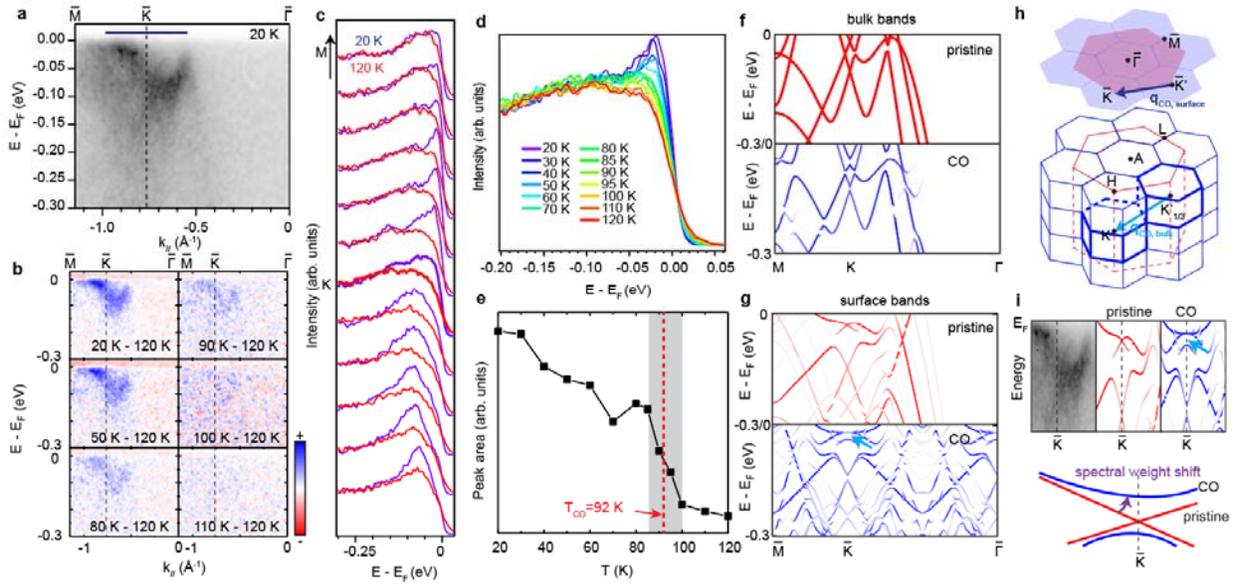

**Figure 6. Temperature-dependent band reconstruction induced by the CO. a** Γ-K-M cut in the vicinity of $E_F$ taken at 20 K. **b** Subtraction between cuts measured at selected temperatures and that at 120 K, with the energy and momentum range same as **a**. The intensity is such that white represents zero, indicating no change. **c** EDC comparison taken at 20 K and 120 K in the region denoted by the dark blue bar in **a**. **d** Stacked EDCs as a function of temperature acquired in the region between K and M. **e** Integrated peak area of each curve in **d** between -0.2 eV and $E_F$, plotted as a function of temperature. Grey shade covers the temperature range where an abrupt drop of peak area occurs. The red dashed line indicates T*. **f** Bulk band structure calculated by DFT, in the pristine phase (upper) and the CO phase (lower). **g** Surface bands in the pristine and the CO phase calculated by DFT. The blue arrow points to the gap opening (see text). **h** Schematic of the surface and bulk BZs. Dark blue and light blue arrows correspond to $q_{CO,surface}$ and $q_{CO,bulk}$, respectively. **i** Direct comparison of ARPES data, pristine surface bands and the surface bands in the CO state, and the schematic showing the origin of the spectral weight shift.

**Acknowledgements**

S.D.W. and G.P. gratefully acknowledge support via the UC Santa Barbara NSF Quantum Foundry funded via the Q-AMASE-i program under award DMR-1906325. Work at Rice is supported by the U.S. Department of Energy (DOE) grant No. DE-SC0021421 (MY) and the Gordon and Betty Moore Foundation's EPiQS Initiative through grant no. GBMF9470 (MY). Work at the University of California, Berkeley and Lawrence Berkeley National Laboratory was funded by the U.S. DOE, Office of Science, Office of Basic Energy Sciences, Materials Sciences and Engineering Division under Contract No. DE-AC02-05CH11231 (Quantum Materials Program KC2202). This material is based in part upon work supported by the National Science Foundation Graduate Research Fellowship Program under Grant No. DGE-2140004. Part of the research described in this work was performed at the Canadian Light Source, a national research facility of the University of Saskatchewan, which is supported by Canada Foundation for Innovation (CFI), the Natural Sciences and Engineering Research Council of Canada (NSERC), the National Research Council (NRC), the Canadian Institutes of Health Research (CIHR), the Government of Saskatchewan, and the